# Using observed bacteria concentration and modeled transit time under an analytical framework to estimate overall removal rate of fecal coliform in an estuary

Jiabi Du[1], Jian Shen[2], Kyeong Park[3], Xin Yu[2], Fei Ye[2], Qubin Qin[2], Jilian Xiong[2], Yu Chen[4]
[1]Applied Ocean Physics & Engineering, Woods Hole Oceanographic Institution, Woods Hole, MA 02543, USA
[2]Virginia Institute of Marine Science, College of William and Mary, Gloucester Point, VA 23062, USA
[3]Department of Marine Sciences, Texas A&M University at Galveston, Galveston, TX 77554, USA
[4]State Key Laboratory of Estuarine and Coastal Research, East China Normal University, Shanghai 200062, China

Correspondence to: Jiabi Du (jdu@whoi.edu; jiabi.du@gmail.com)

Highlights:
- Overall removal rate ($K$) is the key parameter for modeling of fecal coliform bacteria.
- A method to estimate $K$ based on observed spatial distribution of bacteria concentration and modeled transit time is proposed.
- The method converts the $K$ estimation from a temporal problem into a spatial problem.
- Credibility of the method is demonstrated by consistent results from multiple approaches.

**Abstract:** Abundance of fecal coliform (FC) is widely used to indicate the potential presence of pathogens, the No.1 cause of water impairments in the U.S. Despite extensive monitoring efforts, assessing and modeling FC pollution still faces challenges, largely owing to the uncertainties in estimation of overall removal rate ($K$). This study proposes an alternative method to estimate *in situ* $K$ by combining observational data, hydrodynamic simulation, and analytical solution. The method requires the observed spatial distribution of FC concentration along an estuarine channel and the numerically-simulated transit time, and converts the $K$ estimation from a temporal problem into a spatial problem, potentially reducing survey duration, effort, and cost. Application of the method gave an estimation of $K = 0.5$ d$^{-1}$ on average for the Nassawadox Creek in Chesapeake Bay. The numerical and analytical model results with the estimated $K$ agreed well with the observation, demonstrating the credibility of the method.

## 1 Introduction

A growing number of rivers, estuaries, and coastal waters becomes impaired by pathogens (Weiskel et al., 1996; Reeves et al., 2004). Reported by the U.S. Environmental Protection Agency (USEPA), there are 9,874 water bodies in US that are impaired by pathogens in 2018 (USEPA, 2019). Because of their low-concentration and high species diversity, the potential presence of pathogenic microorganisms is usually measured by the abundance of indicator bacteria (Canale et al., 1993). Indicator bacteria are not necessarily pathogenic but are found abundantly in polluted waters where pathogenic organisms are likely to exist (Noble et al., 2003). The levels of indicator bacteria in bathing waters have been shown to correlate with the incidence of illness in swimmers (Haile et al., 1999). Among numerous viruses, bacteria, and protozoa in the polluted waters, fecal coliform (FC) is one of the most commonly used indicators (Servais et al., 2007; Ouattara et al., 2013; de Brauwere et al., 2014). Concentration of FC is regularly monitored to ensure that water bodies meet established sanitary standards for different uses, such as coastal shellfish harvesting, drinking water resources, and health security in recreational waters (Noble et al., 2003; Chigbu et al., 2005; Guber et al., 2006; Gronewold and Wolpert, 2008). Water quality managers seek to regulate FC loads so that water quality standards are not violated. Violations of these standards, however, still frequently occur particularly after intense precipitation events that result in a large quantity of raw sewage and surface runoff discharging into the receiving water body (Auer and Niehaus, 1993).

FC bacteria are introduced to water bodies from two types of sources, diffuse and point sources (Fig. 1). The diffuse sources mainly comprise of the discharge of FC in surface runoff, mostly from manure applied as fertilizer or from grazing animals (Hunter et al., 1999; Crowther et al., 2002; Reder et al., 2015). They also include the direct input to the surface water from waterfowls and the resuspension from streambeds (Valiela et al., 1991; Pachepsky and Sheldon, 2011; Cho et al., 2016). Point sources include discharge from sewage treatment plants, leaks from onsite sewage facilities, stormwater overflow, direct discharge from agriculture and meat processing plants, etc. (Tian et al., 2002; Desai et al., 2011). After being discharged to the receiving water, FC decreases or disappears from the water column with time, primarily through sunlight inactivation, protozoan predation, and settling/sedimentation as well as dilution (Fig. 1; Dan and Stone, 1991; Gannon et al., 2005; Rochelle-Newall et al., 2015). Except for a limited number of occasions, in which the FC can persist or even increase, for example, in a nutrient and organic matter rich, low oxygen tropical soil environment (Ishii and Sadowsky, 2008), the bacterial concentration generally decreases with time.

Extensive modeling studies have been conducted to understand the transport and fate of FC in the water column and provide decision-support information for effective public health management (Collins and Rutherford, 2004; Cho et al., 2016). Various mechanistic water quality models coupled with hydrodynamic models have been developed and applied for predicting the bacterial conditions in different water bodies (e.g., Kashefipour et al., 2002; Steets and Holden, 2003; Liu et al., 2006; Manache et al., 2007; Gao et al., 2011, 2015; Rodrigues et al., 2011; Romeiro et al., 2011; Liu and Huang, 2012; Feng et al., 2013; de Brauwere et al., 2014; Islam et al., 2018). Nevertheless, the uncertainties in bacterial model parameters (notably, the removal rate) as well as in the bacterial loading still greatly hamper the credibility of the modeling results.

The decrease in bacterial concentration ($C$) with time generally follows first-order kinetics (Thomann and Mueller, 1987):

$$\frac{dC}{dt} = -KC \tag{1}$$

where $t$ is time and $K$ is the first-order overall removal rate typically varying between 0.2 and 3.0 $d^{-1}$ (Mancini, 1978; Crane et al., 1980; Mills et al., 1992; Auer and Niehaus, 1993; Chigbu et al., 2005; Liu et al., 2006). A number of biotic and abiotic factors influence $K$, including algal toxins, bacteriophages, nutrient, pH, predation, temperature, salinity, and irradiance (Bowie et al., 1985; Auer and Niehaus, 1993). Among these, irradiance and temperature are generally considered the most important (Auer and Niehaus, 1993; Esham and Sizemore, 1998; Xu et al., 2002; Menton et al., 2003). It is commonly agreed that a higher temperature or irradiance tends to result in a larger removal rate (e.g., Chigbu et al., 2005). The influence of various processes has been studied for $K$. For example, Manache et al. (2007) considered the effects of water temperature, solar radiation, salinity, and sedimentation on the removal rate. Servais et al. (2007) and Liu et al. (2015) added the settling velocity as a parameter to take into account the removal by sedimentation.

The overall removal rate $K$ represents the combined effect of all biogeochemical processes that affect FC. There are a variety of different empirical formulations for $K$, varying from laboratory conditions to natural conditions and from one system to another. One widely used laboratory method is to obtain the time-series of $C$ in a container and estimate $K$ from the rate of decrease in $C$ (e.g., Davies et al., 1995). While the influence of a single factor can be measured by laboratory experiments, the combined effect of various factors (light, salinity, temperature, sedimentation, etc.) and their diurnal or semidiurnal fluctuations in coastal waters are difficult, if not impossible, to include under laboratory condition. Another way to estimate $K$ is based on *in situ* time-series data of $C$ at one location over a relatively long period, e.g., multiple days (Bellair et al., 1977; Chigbu et al., 2005; Jenkins et al., 2011). However, the observed decrease in $C$ can be caused by both biogeochemical processes-induced removal and physical transport-induced dilution. The resulting $K$ is likely to be overestimated without excluding the influence from physical dilution. In modeling FC, the $K$ value has been usually estimated through a trial-and-error approach by adjusting it iteratively to obtain a good agreement between model and observation. Not only is this method time-consuming, it also suffers from the uncertainties in watershed loading. When the loading is overestimated, the resulting $K$ through the trial-and-error approach will also be overestimated. Therefore, an efficient method to estimate $K$ based on *in situ* data would be very valuable for accurate modeling of FC.

In this study, we propose a method to estimate $K$. We first derived an analytical solution to decouple the influence of physical transport and biogeochemical processes, with the former quantified with a transport timescale, transit time. Under the analytical framework, $K$ then can be estimated from the observed spatial distribution of FC concentration and the transit time computed by a hydrodynamic model. This method was applied to the Nassawadox Creek, a sub-estuary with multiple tributaries located in the eastern shore of Chesapeake Bay. A water quality model with the estimated $K$ was applied to model the FC in the estuary. A good agreement between the model and observation demonstrates the validity and rationale of the proposed method of estimating $K$.

## 2 Methods
*2.1 Analytical model*

A simplified analytical solution is used to examine the relative importance of physical transport and biogeochemical processes. The analytical model employs a transport timescale to quantify the influence of physical transport and an overall removal rate to quantify the influence of biogeochemical processes. A simplified one-dimensional governing equation for FC involving advection and first-order removal may be expressed as:

$$\frac{\partial C}{\partial t} + u\frac{\partial C}{\partial x} = -KC + R \quad (2)$$

where $x$ is the along channel distance measured from the headwater ($x = 0$), $u$ is the along-channel velocity, and $R$ is the external load (except the load from the headwater at the upstream end). The terms from left to right in Eq. 2 represent the local change in $C$ with time, advective transport, overall removal, and external sources, respectively. The loading from the headwater is implicitly included in the boundary condition at headwater, $C(x=0) = a$. In a steady state, Eq. 2 has a solution of:

$$C(x) = a \cdot e^{-Kx/u} + \frac{R}{K}\left(1 - e^{-Kx/u}\right) \tag{3}$$

which gives a straightforward interpretation of how physical transport and biogeochemical processes affect the distribution of FC concentration. It indicates that the FC concentration decreases downstream exponentially if not considering the external sources, with the removal rate $K$ and flow speed $u$ being the key parameters that regulate the decreasing trend. A larger removal rate and/or a slower flow will cause a faster decreasing of the FC concentration toward the downstream and vice versa. The solution is a good approximation when considering the tidally averaged condition and has been applied successfully for estuaries and streams (Liu et al., 2008; Shen and Zhao, 2010).

Note that Eq. 2 does not take into account the diffusion process proportional to $\partial^2 C/\partial x^2$, which can sometimes be important. In a still lake or pond, for example, the diffusion is the primary mechanism for the pollutant dispersion. From a Lagrangian perspective, the timescale $x/u$ represents the mean freshwater age at a location $x$ for an ideal regular channel (Zimmerman, 1976; Deleersnijder et al., 2001). Water age, defined as the time elapsed since the water parcel enters the water body through one of the defined boundaries, takes into account both the advection and diffusion processes. More importantly, water age also includes the influence of the return flow as part of water exported during previous ebb tide will re-enter the water body during the following flood tide. With $x/u$ replaced by water age ($\alpha$), the bacterial concentration can be expressed as:

$$C(x) = ae^{-K\alpha} + \frac{R}{K}(1 - e^{-K\alpha}) \tag{4}$$

which gives a straightforward Lagrangian perspective of the change in FC concentration, that is, the concentration changes exponentially with the water age.

Application of Eq. 4 is still not easy as it requires information for the spatio-temporally varying water age. For most estuaries with regular geometry (i.e., no dramatic change in cross-channel width), the water age typically shows a linearly increasing trend toward downstream, e.g., in the Potomac, Rappahannock, York, and James rivers in Chesapeake Bay (Fig. 6 in Shen and Wang, 2007). Instead of using a spatially varying water age, we can further simplify Eq. 4 by using another scalar transport timescale, transit time ($\phi$), defined as the lifetime for riverine material from entering to exiting the water body (Shen and Haas, 2004) and thus equivalent to the water age at the downstream end of the water body, i.e., $\phi \equiv \alpha$ at $x = L$ where $L$ is the length of the system. With the assumption of a linear relationship between water age and distance, i.e., $\alpha = x\phi/L$, Eq. 4 can be transformed into:

$$C(x) = ae^{-Kx\phi/L} + \frac{R}{K}(1 - e^{-Kx\phi/L}) \tag{5}$$

which enables the estimation of the spatial distribution of $C$ based on three scalar parameters, $K$, $\phi$, and $L$. The transit time $\phi$ can be calculated not only from a numerical simulation of freshwater age but can also be estimated by other methods, such as lumped parameter models based on environmental tracer data (Amin and Campana, 1996; McGuire and McDonnell, 2006), biogeochemical methods based on decay of radionuclides (Wallbrink et al., 1998; Cartright and Morgenstern, 2016), and one-dimensional mixing and flushing models (Helder and Ruardij, 1982; Sheldon and Alber, 2002). Using the transit time makes the application of the analytical solution easier and flexible while not compromising its scientific integrity. It is important to note that transit time varies with time due to time varying river discharge, wind, and tide (Wang et al., 2004), which makes numerical simulations a more feasible and economic option.

For many cases, the contribution of the downstream sources is minor when compared to the upstream sources from the headwater due to the larger size of the upstream watershed. Additionally, the influence of downstream sources tends to diminish quickly due to a strong dilution with oceanic water coming into the system from the downstream end and their influence to the upper reach is further reduced due to the downstream-ward surface residual current. FC concentration in natural waters varies over several orders of magnitude and it is reasonable to neglect those minor downstream sources for a quick environmental assessment. In such cases, the concentration of FC is mainly controlled by the upstream sources from the headwater. Then, Eq. 5 can be reduced to:

$$C(x) = ae^{-Kx\phi/L} \tag{6}$$

which shows that the distribution of $C$ depends on the non-dimensional number $\phi K$ that represents the influence of physical transport ($\phi$) and biogeochemical processes ($K$). A larger $K$ for a given $\phi$ or a larger $\phi$ for a given $K$ will result in a faster decreasing of FC toward downstream (Fig. 2). A larger $K$ certainly removes more FC and a larger $\phi$ means slower water flow leaving longer time for bacterial removal. For any given estuary or a segment of it, it is the relative variability between $\phi$ (transport) and $K$ (biogeochemical removal) that determines the dominant process. For instance, the relative magnitude of the variability in $K$ and $\phi$, represented by the standard deviation ($\sigma$) normalized by its mean (overbar), can be used to indicate the dominant process:

$$\frac{\sigma(K)}{\overline{K}} \gg \frac{\sigma(\phi)}{\overline{\phi}}, \text{ removal dominant} \tag{7}$$

$$\frac{\sigma(K)}{\overline{K}} \ll \frac{\sigma(\phi)}{\overline{\phi}}, \text{ transport dominant} \tag{8}$$

While $K$ is known to be influenced by temperature, radiation, and sedimentary processes, the variability of $\phi$ depends on a number of factors, including tide, freshwater, wind, bathymetry, and geometry. A smaller and shallower estuary where the flushing capacity is more sensitive to changes in external forcing tends to have higher variability in $\phi$, in contrast to a larger and deeper estuary that typically shows relatively little variability in $\phi$.

It is worthy to note that the analytical solution can be also applied for other pollutants (e.g., nutrient). Nutrient pollution is very common for coastal systems; excessive nutrient loading results in eutrophication, causing a series of adverse impact to coastal ecosystems (Kemp et al., 2005; Howarth, 2008). The observed total nitrogen also shows an exponentially decreasing trend toward downstream in Chesapeake Bay (Du and Shen, 2017). One big difference between nitrogen and FC is that the overall nitrogen removal rate is much smaller, on the order of 0.01 d$^{-1}$ for Chesapeake Bay (Dettmann, 2001).

*2.2 Estimation of removal rate*

One application of the analytical model is to inversely estimate $K$ (overall removal rate) by using the observed $C$ and modeled $\phi$. The removal rate can be calculated using Eq. 6 with data from two locations. With the concentrations of $C_1$ and $C_2$ at locations $x_1$ and $x_2$, respectively, Eq. 6 gives:

$$\ln(C_1) = \ln(a) - \frac{K\phi x_1}{L} \tag{9}$$

$$\ln(C_2) = \ln(a) - \frac{K\phi x_2}{L} \tag{10}$$

Subtracting Eq. 10 from Eq. 9 to eliminate the unknown boundary value *a* gives:

$$\ln(C_1) - \ln(C_2) = -K\phi \frac{(x_1 - x_2)}{L} \tag{11}$$

with which we can estimate *K* from the slope of ln(*C*) against the distance *x*:

$$K = \frac{-\beta L}{\phi} \tag{12}$$

where the slope $\beta = (\ln(C_1) - \ln(C_2)) / (x_1 - x_2)$. When data from more than two locations are available, the slope $\beta$ can be estimated using linear regression.

*2.3 Study site and observation*

The proposed method to estimate *K* was applied to the Nassawadox Creek, located in the Northampton County, Virginia, on the west side of the Eastern Shore of the Delmarva Peninsula (Fig. 3a). The estuary has a watershed of 76.1 km$^2$ and drains westward to Chesapeake Bay, the largest estuary in the U.S. It is 8 km long from its mouth to the head, and its width changes from 300 m in the upper part to 900 m in the lower part. Several smaller creeks join from both sides, forming a small estuarine system. Dominant land uses in the Nassawadox Creek watershed include agriculture (50%), forest (20%), and wetland (18%), which combine to account for 88% of the entire watershed. The Nassawadox Creek experiences average temperatures of 2-9°C in January and 22-30°C in July. Average annual precipitation is 1049 mm (41.3 inches). In the middle portion of the Creek, most of the salinities range from 15-25 psu. Due to the short length and limited freshwater input (0.75 m$^3$ s$^{-1}$ on average), the salinity gradient from mouth to head is small.

The Nassawadox Creek is one of important nursery habitats for shellfish. Frequent occurrences of FC pollution caused economic loss and raised health concerns from local communities. To monitor the FC pollution, the Virginia Department of Environmental Quality (DEQ) has conducted continuous nearly-monthly measurements of FC concentration inside the creek. During 2007-2012, 70 surveys were conducted to collect data for salinity at 4 stations and FC concentration at 46 stations throughout the creek by the Virginia DEQ (Fig. 3b).

*2.4 Watershed model*

A watershed model, LSPC (Loading Simulation Program C$^{++}$) (Shen et al., 2005a), was applied to estimate the freshwater and FC loadings into the creek. LSPC is a stand-alone personal computer-based watershed modeling system developed in Microsoft C$^{++}$ that integrates GIS tools, data management capabilities, a post-processor, and a dynamic watershed model within a Windows environment in which the simulation algorithms of hydrology and water quality are based on the Hydrologic Simulation Program - FORTRAN (HSPF). Like other watershed models, LSPC is a precipitation-driven model and requires necessary meteorological data as model input. LSPC has been applied to estimate the bacterial loading into the Nassawadox Creek and the model show good performance in reproducing the freshwater discharge (Shen et al., 2005b). This LSPC application was used in this study to estimate the freshwater and FC loadings into the creek.

*2.5 Hydrodynamic model*

Physical transport, as simulated by a hydrodynamic model, is an essential process in driving the movement of dissolved and particulate substances in aquatic systems. We used in this study the hydrodynamic model, the Environmental Fluid Dynamics Computer Code (EFDC) (Hamrick, 1992). EFDC is a general-purpose modeling package for simulating 1-, 2-, and 3-dimensional flow and transport in surface water systems including rivers, lakes, estuaries, reservoirs, wetlands, and coastal oceans. It was originally developed at the Virginia Institute of Marine Science for estuarine and coastal applications and is a public domain software. The model has been extensively tested and documented. The EFDC model has been integrated into the EPA's Modeling Toolbox for supporting water quality management (http://www.epa.gov/exposure-assessment-models/efdc, last accessed on September 27, 2019).

The model domain covers the entire Nassawadox Creek and its tributaries with the open boundary extended into Chesapeake Bay (Fig. 3a). Setting the model boundary well outside the area of interest enhances the model accuracy by reducing the influence of computational noise at the boundary. A curvilinear orthogonal grid with 3364 surface cells was generated to align the grid edge with the shoreline (Fig. 3a). The grid resolution ranges from 200 m at the open boundary to 20 m inside the creek, with 3 sigma layers vertically. Since the system is shallow and usually well mixed due to limited freshwater discharge, there is no necessity of using a finer vertical resolution. The model bathymetry was estimated based on the 3-arc-second resolution Coastal Relief Model (https://www.ngdc.noaa.gov/mgg/coastal/crm.html, last accessed on September 27, 2019). The model was driven by open boundary conditions for water level and salinity from the output of the validated EFDC Chesapeake Bay model (Hong and Shen, 2012; Du and Shen, 2015, 2016) and atmospheric forcing from the National Centers for Environmental Prediction. The hydrodynamic model was validated using the salinity data in 2007-2012 collected by the Virginia DEQ. The age of freshwater from the headwater of each tributary was computed following the method proposed by Deleersnijder et al. (2001).

*2.6 Water quality model for fecal coliform*

A water quality model for FC concentration coupled with the estimated *K* (Section 2.2) was used to assess the credibility of the proposed method of *K* estimation. The governing advection-diffusion mass-balance equation for FC concentration with the mortality of FC parameterized by an overall removal rate can be expressed as:

$$\frac{\partial C}{\partial t} + \frac{\partial}{\partial x}(uC) + \frac{\partial}{\partial y}(vC) + \frac{\partial}{\partial z}(wC) = \frac{\partial}{\partial x}\left(A_x \frac{\partial C}{\partial x}\right) + \frac{\partial}{\partial y}\left(A_y \frac{\partial C}{\partial y}\right) + \frac{\partial}{\partial z}\left(A_z \frac{\partial C}{\partial z}\right) + S_C - KC \quad (13)$$

where ($u$, $v$, $w$) and ($A_x$, $A_y$, $A_z$) are velocities and turbulent diffusivities in *x*-, *y*- and *z*-directions, respectively, and $S_C$ is external sources. The physical transport processes, represented by the advective (last three terms on the left-hand side) and turbulent diffusive (first three terms on the right-hand side) transport, are identical to those in the mass-balance equation for salinity in the hydrodynamic model (Hamrick, 1992). The last term in Eq. 13 represents the overall removal of FC by various biogeochemical processes. In the EFDC model, Eq. 13 is solved using a fractional step procedure that decouples the kinetic terms from the physical transport terms (Park and Kuo, 1996; Tetra Tech, 2007). The solution schemes for both the physical transport and the kinetic equations are second-order accurate. The water quality model results were compared to the observed FC concentration data in 2007-2012 collected by the Virginia DEQ.

To examine the relative importance between physical transport and biogeochemical (FC removal) processes, we conducted two sensitivity runs with the same model configuration and loading but with *K* of 0 and 0.25 d$^{-1}$, respectively. The simulated distribution of FC with *K* = 0 is determined by the dilution (physical transport) only and thus the difference from the base run will indicate the influence of the FC removal. The difference between the base run and the run with $K = 0.25$ d$^{-1}$ will indicate the influence of reduced FC removal.

## 3 Results
*3.1 Observation*

The observed distribution of FC was characterized by an evident decreasing trend from upstream to downstream (Fig. 3b), with high concentrations at and near the headwaters of the Creek and tributaries and much lower in the lower-middle part. The mean FC concentration at all stations in the mainstem was less than 15 cfu (colony forming unit) per 100 ml, suggesting an efficient dilution induced by exchange with the low-FC bay water. Importantly, the median values of FC concentration decreased toward downstream in a nearly exponential function of the distance for the three main tributaries (Fig. 4), which validates the fundamental assumption for the proposed method of $K$ estimation (Eq. 6).

*3.2 Hydrodynamic model*

The hydrodynamic model simulates well the variation of observed salinity at four monitoring stations (Fig. 5). We used two indexes, model skill and root mean square error (*RMSE*), for quantitative assessment of the model performance. The model skill, following Willmott (1981), provides an index of model–observation agreement, with a skill of 1 indicating perfect agreement and a skill of 0 indicating complete disagreement. The *RMSE* indicates the magnitude of average deviation between model and observations. The model skills are larger than 0.70 and the *RMSE* values are in the range of 2-3 psu (Fig. 5). The overall temporal variation in salinity was quite consistent among the four stations. It is worthy noting that salinity at the downstream near the mouth had a strong tidal signal (Fig. 5e) while salinity at the upstream stations had a relatively weak one. The age of freshwater from the headwater computed with the hydrodynamic model following the method in Deleersnijder et al. (2001) shows water age increasing almost linearly with distance toward the downstream; Fig. 6 shows Tributary #1 as an example. It hence justifies the simplification procedure for Eq. 5 of using one scalar timescale, transit time, to quantify the overall influence of physical transport along the entire channel.

*3.3 K estimation*

The removal rate was estimated based on the observed exponentially decreasing *C* along the channel of each of the three main tributaries (see Fig. 3b for their locations). Only the survey data exhibiting a significant linear trend (*p*-value < 0.05) between ln(*C*) and *x* were used for the estimation. Over the 70 surveys, 17, 13, and 11 surveys were used for Tributary #1 to #3, respectively. The median values of the estimated removal rates vary among tributaries, with a range of 0.35-0.6 d$^{-1}$ and a mean of 0.5 d$^{-1}$ (Fig. 7).

*3.4 Water quality model: validation of the K estimation*

With the overall removal rate 0.5 d$^{-1}$, the water quality model gives a good reproduction of the observations (Figs. 8). The variability of modeled FC is very high because the FC loading is driven by the episodic surface run-off (or precipitation) events. Different from freshwater discharge, which always has a background value, the diffuse FC loading exists only after precipitation during which surface run-off erodes the accumulated animal feces on land. Both the model results and observations consistently show relatively high concentration of FC in the tributaries (Fig. 8b,d), with values larger than 100 cfu per 100 ml, while the FC concentration is small and even undetectable near the Creek mouth (Fig. 8a) where water is quickly diluted by the outside low-FC bay water. In terms of the long-term mean as well as the distribution of cumulative frequency, the water quality model results reproduces the observations well (Fig. 9). We acknowledge that the model tends to underestimate the long-term mean values, particularly at stations with higher FC concentration (Fig. 9b). The bias could be caused by the underestimated FC loading from the watershed model or the complex geometry in the upstream where the model grid resolution is not fine enough. In addition, there is also a good agreement between the numerical and analytical models for the 6-year mean FC distribution in the three tributaries (Fig. 10). The good agreement suggests not only the rationale of the removal rate estimation but also the potential credibility of the analytical model.

Two sensitivity runs with regard to $K$ show that the FC concentration increases throughout the system with no bacterial removal ($K = 0$). Compared to the base run ($K = 0.5$ d$^{-1}$), the spatial distribution of FC remains similar but the concentration becomes much higher with the difference increasing toward downstream of Tributary #2 (Fig. 11). The numerical experiments show that the removal reduces the FC concentration by as large as 95%. The bacterial removal becomes more effective toward downstream because of longer water age in the downstream part and the dilution with cleaner water from the downstream end. The same spatial trend with or without bacteria removal suggests that the spatial distribution is mainly controlled by the physical transport. On the other hand, the bacteria removal plays a key role in determining the severity of FC pollution.

**4 Discussion**

By combining observational data and numerical simulation, we propose a method that can reasonably estimate *in situ* overall removal rate of FC. The method has several advantages, compared to conventional methods based on temporal decreases of FC concentration. First, it converts the temporal problem (i.e., a time-series of FC concentration) into a spatial problem (i.e., spatial distribution data), greatly reducing the measurement duration and hence the cost. Second, it can be used to estimate the *in situ K*, as it takes into account the physical influence. The *in situ K* includes the influence of physical and biogeochemical processes, making the estimated $K$ readily applicable to the given system. Third, with the estimated $K$, the efficiency of a numerical model in assessing the FC pollution would be greatly enhanced as it does not involve the time-consuming trial-error procedure of determining $K$.

It is worth noting that the method may not be suitable and lead to biased estimation in some cases. Some assumptions are adopted when deriving Eq. 6: (1) the downstream sources have negligible influence; and (2) the upstream sources through headwaters have reached or influenced the downstream stations. If the assumptions are not satisfied, the removal rate will be over or under-estimated. For instance, if the sampling is conducted before the high FC water reaches the downstream stations, the $K$ value could be overestimated if data from all stations are used (Fig. 12b). An alternative strategy in this case would be to estimate the decreasing trend for each segment (e.g., estimate the slope separately for each of two segments in Fig. 12b). If there is a significant source in the downstream part, the $K$ value could be underestimated if data from all stations are used (Fig. 12c).

Additionally, the transit time may vary with time particularly for regions receiving highly variable atmospheric forcings, freshwater input, or the adjacent shelf conditions. To obtain the temporally varying transit time, a simulation of the hydrodynamic condition and water age (transit time) is needed for the target period. The transit time is usually regulated by freshwater discharge (Wang et al., 2004). The relationship between freshwater and transit time can be easily obtained by conducting sensitivity tests using a numerical model (e.g., Shen and Haas, 2004). The relationship, together with freshwater input for a given period, can then be used for a quick estimation of the transit time. When the water age does not change linearly with the distance, one could use the original formula (Eq. 5) instead of Eq. 6 to estimate the removal rate by a trial-and-error procedure, that is, simply using different removal rate to find the appropriate one that give the best match of estimated FC concentration against observed values.

## 5 Conclusions

This study highlights the robustness of the proposed method to estimate the overall removal rate of FC by utilizing the spatial distribution of observed FC concentration and a hydrodynamic model. Consistent comparison among observation and numerical and analytical model results demonstrates the credibility of the method. The method reduces measurement duration and cost by utilizing spatial distribution data. The estimated *in situ K* includes the influence of physical and biogeochemical processes, making it readily applicable to a target estuary. The method provides estimates of $K$, without the time-consuming trial-error procedure of determining $K$, thus enhancing the efficiency of numerical model application in assessing the FC pollution.

With a warming climate, more intensified precipitation events, and changing watershed properties (e.g., more impervious land by urbanization), FC pollution is likely to pose an increasing threat to the quality of recreational and fishery harvesting waters (Rochelle-Newall et al., 2015). Efficient assessment of the potential extent of FC pollution is thus of urgent need for fast management response. The method proposed in this study will likely enhance the efficiency of FC pollution assessment.

## Acknowledgments

This project was supported by the Virginia Department of Environmental Quality (Contract #16006). We thank Anna Schlegel and Kristie Britt for providing data and many suggestions for this study.## References

Amin, I. E., Campana, M. E. (1996). A general lumped parameter model for the interpretation of tracer data and transit time calculation in hydrologic systems. *Journal of Hydrology*, *179*, 1–21. https://doi.org/10.1016/0022-1694(95)02880-3

Auer, M. T., Niehaus, S. L. (1993). Modeling fecal coliform bacteria-I. Field and laboratory determination of loss kinetics. *Water Research*, *27*, 693–701. https://doi.org/10.1016/0043-1354(93)90179-L

Bellair, J. T., Parr-Smith, G. A., Wallis, I. G. (1977). Significance of diurnal variations in fecal coliform die-off rates in the design of ocean outfalls. *Journal (Water Pollution Control Federation)*, *49*(9), 2022–2030. https://www.jstor.org/stable/25039395

Bowie G. L., Mills W. B., Porcella D. B., Campbell C. L., Pagenkopf J. R., Rupp G. L., Johnson K. M., Chan P. W. H., Gherini S. A., Chamberlain C.E. (1985) Rates, constants, and kinetics formulations in surface water quality modeling (2$^{nd}$ edition). EPA/600/3-85/040, Environmental Research Laboratory, U.S. Environmental Protection Agency, Athens, GA.10

Ishii, S., Sadowsky, M. J. (2008). *Escherichia coli* in the environment: Implications for water quality and human health. *Microbes and Environments*, *23*, 101–108. https://doi.org/10.1264/jsme2.23.101

Islam, M. M. M., Sokolova, E., Hofstra, N. (2018). Modelling of river faecal indicator bacteria dynamics as a basis for faecal contamination reduction. *Journal of Hydrology*, *563*(June), 1000–1008. https://doi.org/10.1016/j.jhydrol.2018.06.077

Jenkins, M. B., Fisher, D. S., Endale, D. M., Adams, P. (2011). Comparative die-off of *Escherichia coli* 0157:H7 and fecal indicator bacteria in pond water. *Environmental Science and Technology*, *45*(5), 1853–1858. https://doi.org/10.1021/es1032019

Kashefipour, S. M., Lin, B., Harris, E., Falconer, R. A. (2002). Hydro-environmental modelling for bathing water compliance of an estuarine basin. *Water Research*, *36*, 1854–1868. https://doi.org/10.1016/S0043-1354(01)00396-7

Kemp, W. M., Boynton, W. R., Adolf, J. E., Boesch, D. F., Boicourt, W. C., Brush, G., Cornwell, J. C., Fisher, T. R., Glibert, P. M., Hagy, J. D., Harding, L. W., Houde, E. D., Kimmel, D. G., Miller, W. D., Newell, R. I. E., Roman, M. R., Smith, E. M., Stevenson, J. C. (2005). Eutrophication of Chesapeake Bay: Historical trends and ecological interactions. *Marine Ecology Progress Series*, *303*, 1–29. https://doi.org/10.3354/meps303001

Liu, L., Phanikumar, M. S., Molloy, S. L., Whitman, R. L., Shively, D. A., Nevers, M. B., Schwab, D. J., Rose, J. B. (2006). Modeling the transport and inactivation of *E. coli* and Enterococci in the near-shore region of Lake Michigan. *Environmental Science and Technology*, *40*, 5022–5028. https://doi.org/10.1021/es060438k

Liu, W.-C., Huang, W.-C. (2012). Modeling the transport and distribution of fecal coliform in a tidal estuary. *Science of the Total Environment*, *431*, 1–8. https://doi.org/10.1016/j.scitotenv.2012.05.016

Liu, W.-C., Chan, W.-T., Young, C.-C. (2015). Modeling fecal coliform contamination in a tidal Danshuei River estuarine system. *Science of the Total Environment*, *502*, 632–640. https://doi.org/10.1016/j.scitotenv.2014.09.065

Liu, Y., Yang, P., Hu, C., Guo, H., (2008). Water quality modeling for load reduction under uncertainty: A Bayesian approach. *Water Research*, *42*, 3305–3314. https://doi.org/10.1016/j.watres.2008.04.007

Manache, G., Melching, C. S., Lanyon, R. (2007). Calibration of a continuous simulation fecal coliform model based on historical data analysis. *Journal of Environmental Engineering*, *133*, 681–691. https://doi.org/10.1061/(ASCE)0733-9372(2007)133:7(681)

Mancini, J. L. (1978). Numerical estimates of coliform mortality rates under various conditions. *Journal (Water Pollution Control Federation)*, 50(11), 2477–2484. www.jstor.org/stable/25040179

McGuire, K. J., McDonnell, J. J. (2006). A review and evaluation of catchment transit time modeling. *Journal of Hydrology*, *330*, 543–563. https://doi.org/10.1016/j.jhydrol.2006.04.020

Menon, P., Billen, G., Servais, P. (2003). Mortality rates of autochthonous and fecal bacteria in natural aquatic ecosystems. *Water Research*, *37*, 4151–4158. https://doi.org/10.1016/S0043-1354(03)00349-X

Mills, S. W., Alabaster, G. P., Mara, D. D., Pearson, H. W., Thitai, W. N. (1992). Efficiency of faecal bacterial removal in waste stabilization ponds in Kenya. *Water Science and Technology*, 26, 1739-1748. https://doi.org/10.2166/wst.1992.0617
13

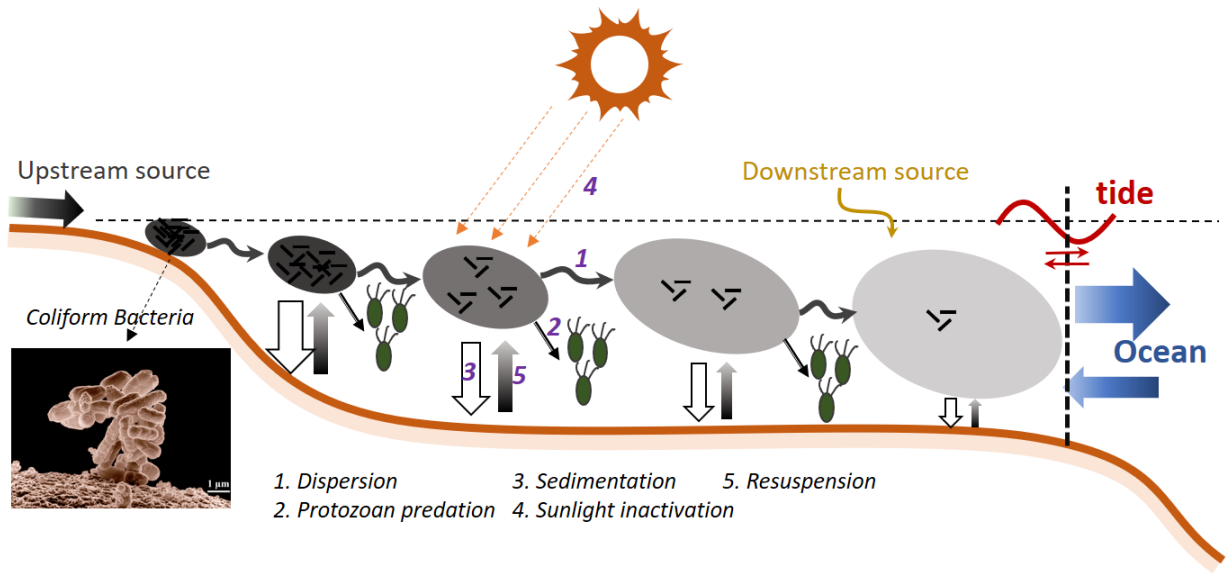

**Figure 1**: A sketch diagram showing the major sources and sinks as well as the transport and exchange processes for the fecal coliform bacteria in an estuary, with the concentration denoted by the filled color in circles (the darker, the higher). This diagram shows how the fecal coliform concentration decreases toward downstream by the dilution and removal processes. Major removal processes include protozoan predation, sedimentation, and sunlight inactivation. The image of the fecal coliform in the lower left corner was released by the Agricultural Research Service of the U.S. Department of Agriculture.

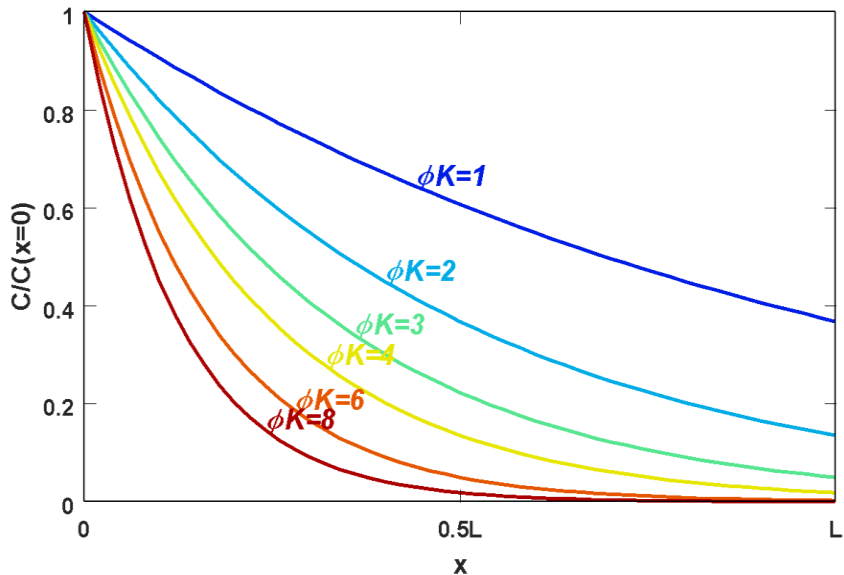

**Figure 2**: Distribution of FC concentration as a function of distance and non-dimensional number $\phi K$ where $\phi$ is the transit time (d), $K$ is the removal rate (d$^{-1}$), and $L$ is the channel length.

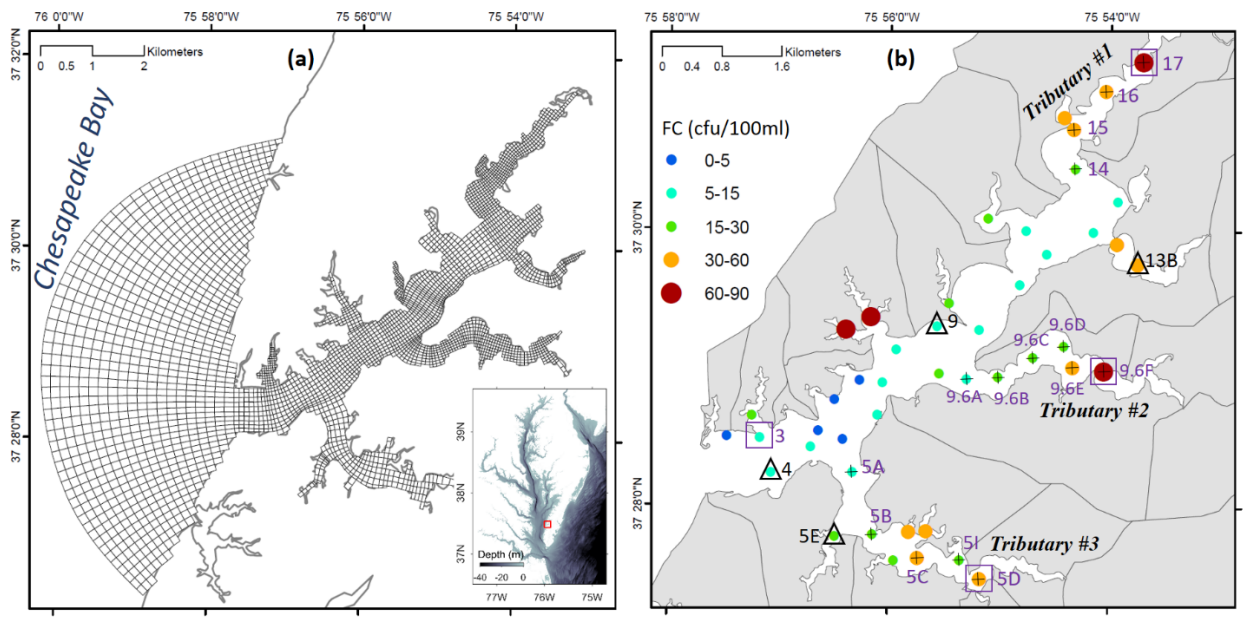

**Figure 3**: (a) Model grid, with the inset showing the location of the Nassawadox Creek in the Chesapeake Bay estuarine system and (b) a map showing the locations of the 46 fecal coliform monitoring stations in the Nassawadox Creek with the color of dots denoting the mean concentration of fecal coliform. Marked in (b) are four stations for salinity used to validate the hydrodynamic model (rectangles), four stations for model-observation FC comparison (triangles), and along-channel stations for Tributary #1 to #3 for the estimation of removal rate (crosses).

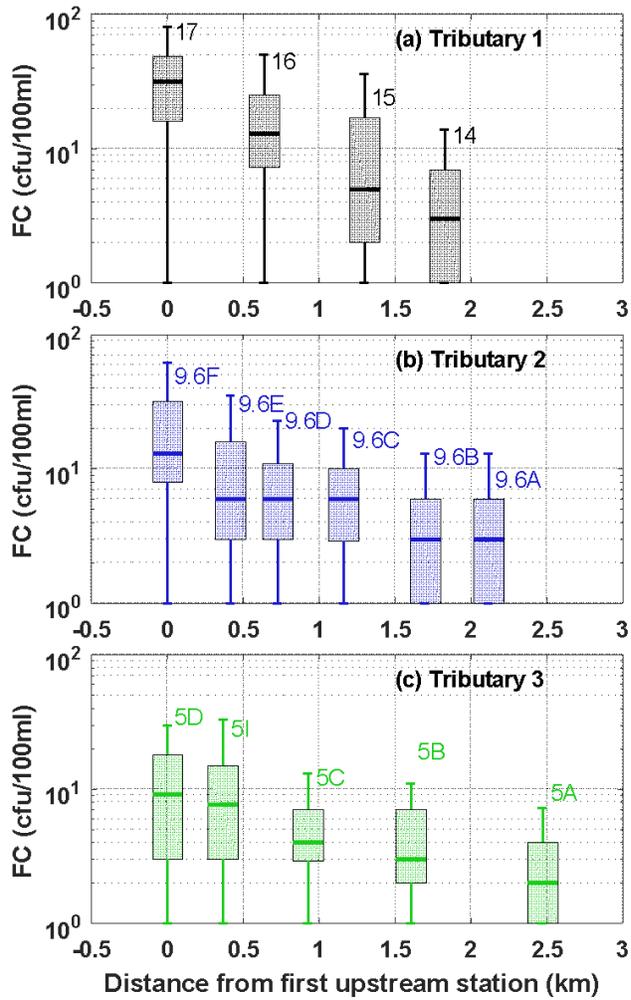

**Figure 4**: Box plots for the observed FC concentration along the channels of Tributary #1 to #3. Note the logarithmic scale for FC concentration (*y*-axis).

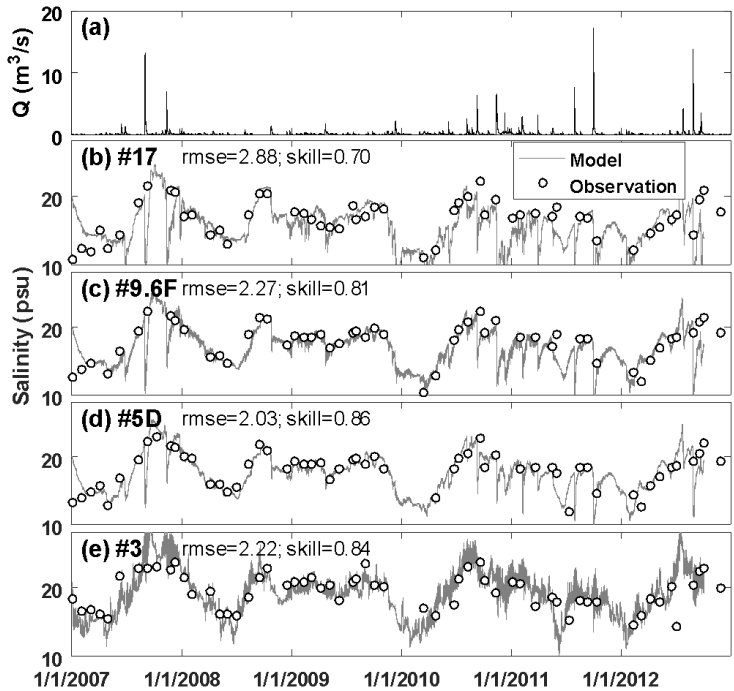

**Figure 5**: (a) Freshwater discharge from the headwater of Tributary #1 and (b-e) comparison of salinity between model and observation at four monitoring stations (see Fig. 2b for their locations).

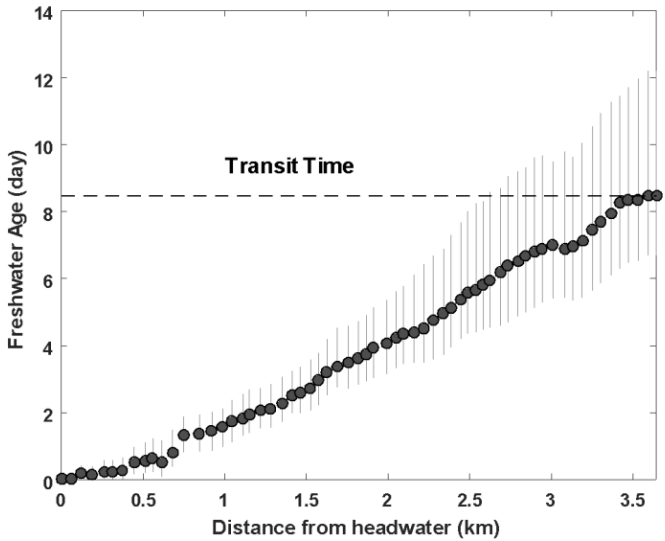

**Figure 6**: Along-channel distribution of freshwater age for Tributary #1 showing median values (solid circles) and the $25^{th}$-$75^{th}$ percentiles (vertical lines) and the transit time (horizontal dashed line). Note the linear trend of water age with the distance from headwater.

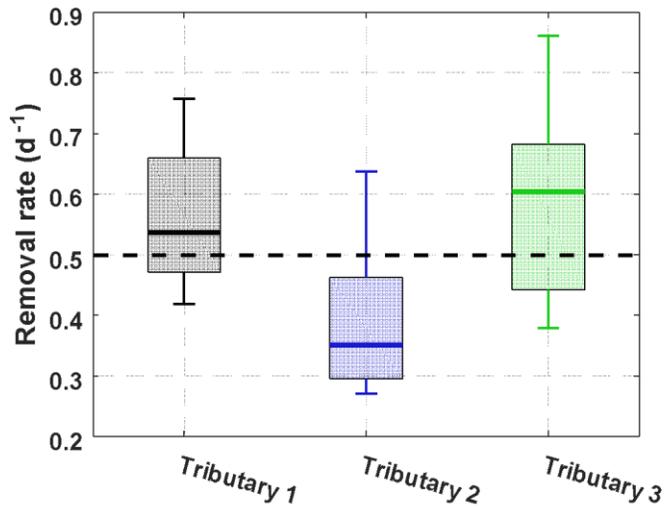

**Figure 7**: Box plot of estimated removal rate for Tributary #1 to #3. The horizontal dashed line indicates 0.5 d$^{-1}$, the mean of median values from three tributaries.

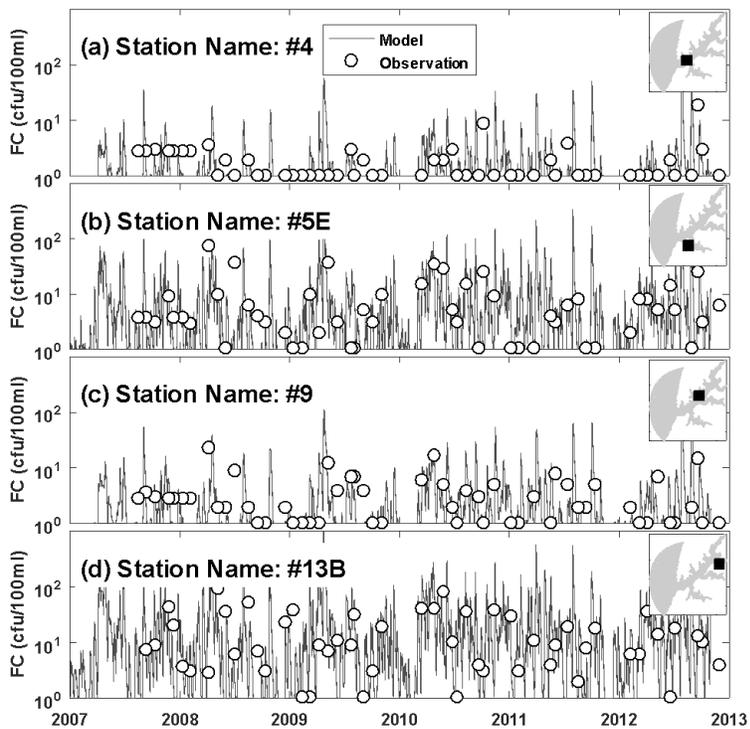

**Figure 8**: Comparison between modeled and observed FC concentration at four selected monitoring stations, with the top-right insets showing station locations (see Fig. 3b for exact locations).

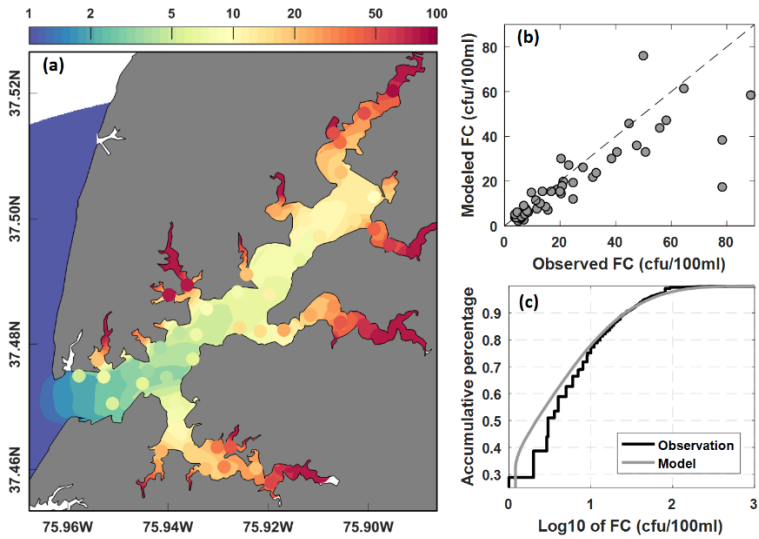

**Figure 9**: Comparison between modeled and observed FC concentration (cfu per 100 ml): (a) 6-year (2007-2012) mean of model results (color contours) and observed data at 46 stations (solid circles), (b) one-to-one comparison of 6-year mean values for 46 stations, and (c) cumulative percentage comparison between model (daily mean values from 46 stations) and observation (a total of 3206 data points from 46 stations).

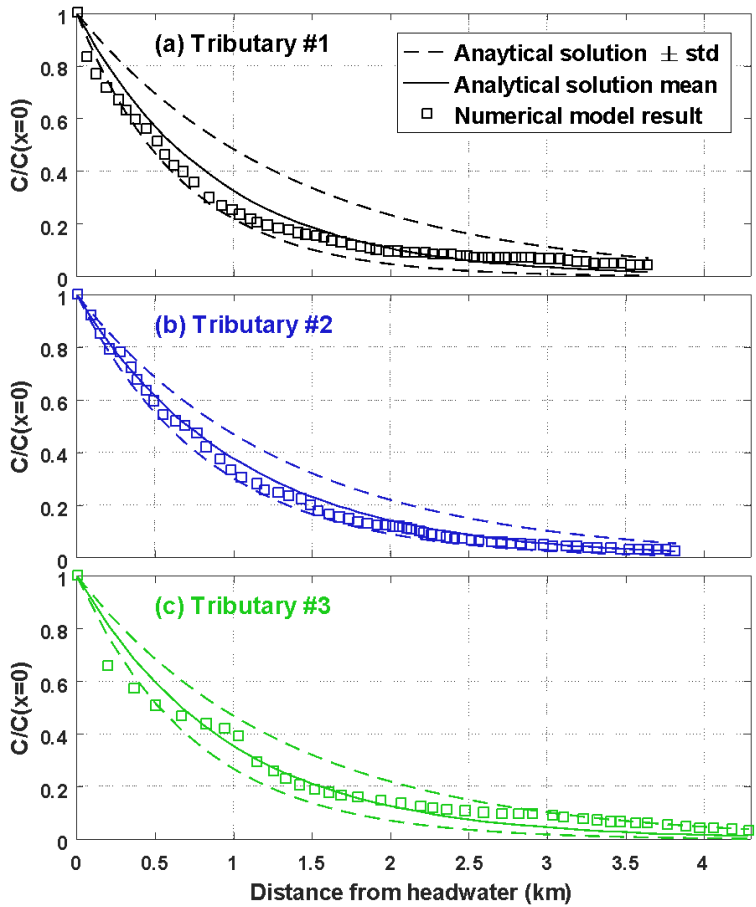

**Figure 10**: Comparison of 6-year mean FC distributions between numerical model (squares) and analytical model with the mean (solid lines) ± one standard deviation (dashed lines) of transit times using $K = 0.5$ d$^{-1}$.

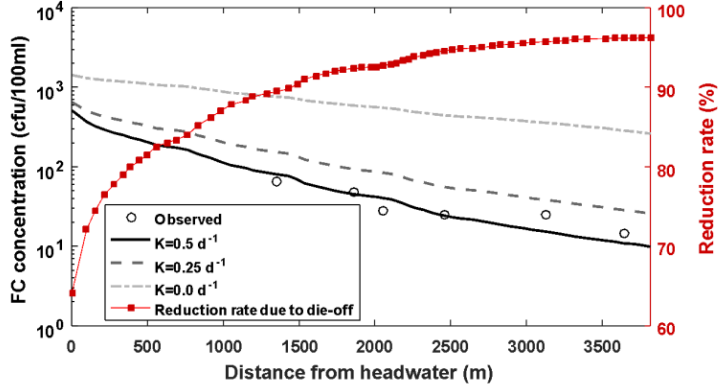

**Figure 11**: The 6-year mean FC distribution along Tributary #2 (see Fig. 2b for its location) from the observations (○) and three model runs with $K = 0.5$, $0.25$ and $0$ $d^{-1}$. The line with ■ indicates the reduction rate (%) due to biogeochemical removal of FC, i.e., the difference between the model results with $K = 0$ and $0.5$ $d^{-1}$. Note the logarithmic scale for FC concentration (left *y*-axis).

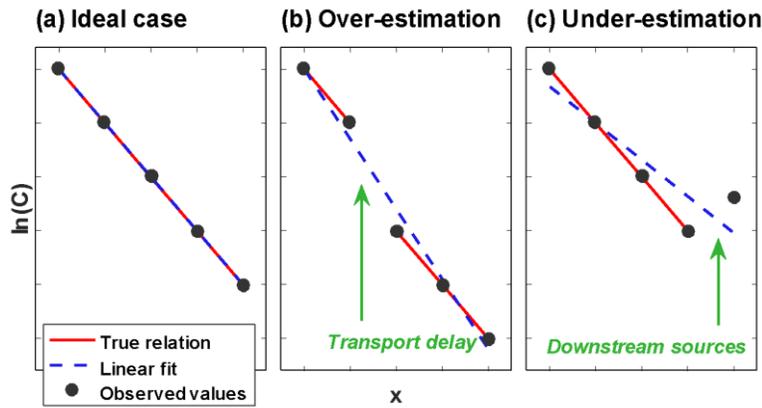

**Figure 12**: Diagrams showing the potential bias in the estimation of the removal rate, with (a) the perfect case with no bias between true rate and estimated rate using a linear regression, (b) an over-estimation due to transport delay, i.e., the upstream sources have not reached the downstream, and (c) an underestimation due to downstream sources that elevate the concentration in the downstream.

23